# Double Splay Nematic Order in Confined Polar Fluids

Zhongjie Ma[1], Miao Jiang[1], Aile Sun[1], Shengzhu Yi[1,2], Jidan Yang[3], Mingjun Huang[3,4], Satoshi Aya[3,4], Qi-Huo Wei[1,5*]

[1] State Key Laboratory of Quantum Functional Materials, Department of Mechanical and Energy Engineering, Southern University of Science and Technology, Shenzhen, 518055, China

[2] Department of Physics, The Hongkong University of Science and Technology, Clear Water Bay, Hong Kong, China

[3] South China Advanced Institute for Soft Matter Science and Technology (AISMST), School of Emergent Soft Matter, South China University of Technology, Guangzhou 510640, China

[4] Guangdong Provincial Key Laboratory of Functional and Intelligent Hybrid Materials and Devices, South China University of Technology, Guangzhou 510640, China

[5] Center for Complex Flows and Soft Matter Research, Southern University of Science and Technology, Shenzhen 518055, China

We report the structure of a modulated phase in the ferroelectric nematic material RM734 under cationic-polymer-coated planar confinement. This phase emerges continuously from the nematic phase upon cooling and transforms into the ferroelectric nematic phase via a first-order transition. Compelling optical evidence, including thickness-dependent periodicity, a sinusoidal variation of the orientation angle, and periodic variations in phase retardance and image intensity, demonstrates that this modulated phase is the theoretically predicted double-splay nematic phase. In this phase, splay deformations alternate along two orthogonal directions oriented at 45° to the substrate surfaces, resulting in a maximum modulation wavelength twice the cell thickness. Our experiments with various ionic coatings reveal that only cationic polymer coatings effectively promote this phase configuration, highlighting the crucial role of electrical screening. These findings further provide insights into its distinctive topological defects within confined geometries.

Nematic (N) liquid crystals exhibit orientational order of anisotropic molecules along a director $\mathbf{n}$, while maintaining inversion symmetry ($\mathbf{n} \equiv -\mathbf{n}$) [1,2]. Ferroelectric nematic ($N_F$) liquid crystals, predicted over a century ago but only recently realized [3-8], comprise of polar, rod-shaped molecules with large electric dipoles. The $N_F$ phase exhibits spontaneous electrical polarization $\mathbf{P}$ [6-9], typically aligned with the director ($\mathbf{P} = P_0\mathbf{n}$). This inversion-symmetry breaking, coupled with flexoelectric and electrostatic interactions, profoundly alters the director field, leading to complex ground states and topological excitations [10-20].

Flexoelectricity arises from the coupling between polarization and elastic deformations in the director field [21,22]. In $N_F$ liquid crystals composed of pear-shaped molecules, splay deformations are energetically favored to achieve close packing [8,23-25], resulting in a positive flexoelectric coupling coefficient ($\lambda$) and an associated free energy: $F_{\text{Flexo}} = -\lambda \int dr^3 \mathbf{P} \cdot \mathbf{n} \nabla \cdot \mathbf{n}$ [15,26,27]. To fill the space, the splay direction can alternate along one axis, leading to the formation of a singlesplay nematic ($N_S$) phase [28-30]. Recent theoretical work by Rosseto and Selinger (hereafter, RS theory) proposes that a double-splay nematic phase, in which splay deformations alternate along two orthogonal axes, may exhibit greater stability than its single splay counterpart [31]. While previous studies have explored one-dimensional modulated textures arising from field-driven flexoelectric instabilities in apolar liquid crystals [32,33], the possibility of two-dimensional modulations has been largely overlooked.

Experimental studies employing precision calorimetry, optical microscopy, and X-ray diffraction have revealed a modulated antiferroelectric phase between the nematic and ferroelectric nematic phases in pure, ion-doped, and mixed samples of the ferroelectric materials RM734 and DIO [34-36]. Although these intermediate phases have been suggested to be splay nematics, the only structural evidence to date is the absence of higher-order X-ray diffraction peaks in DIO which may originate from a sinusoidal director variation or a single splay nematic phase [37]. The precise director configuration of these intermediate antiferroelectric phases—single splay, double-splay [31,38], or otherwise [39]—remains unclear due to a scarcity of direct structural data. Consequently, diverse terminologies, including Ns, Nx, and SmZ$_A$, have been used to describe these phases [24,37,38,40,41].

Electrostatic interactions contribute to the free energy of ferroelectric nematic materials through the following term: $F_{\text{Elec}} = \iint d^3r_1 d^3r_2 \rho(\mathbf{r}_1)\rho(\mathbf{r}_2)e^{-r_{12}/\Lambda}/4\pi r_{12}$. Here $\Lambda$ is the Debye screening length, $r_{12}$ is the distance between the positions $\mathbf{r}_1$ and $\mathbf{r}_2$ [26], and $\rho$ represents the total charge density, encompassing both surface and bound charges. Surface charges originate from the polarization tilt at the confining surfaces, with a density given by: $\rho_S = (\mathbf{P} \cdot \mathbf{v}) \delta(\mathbf{v} \cdot \mathbf{r} - r_{\text{Surf}})$, where $\mathbf{v}$ is the surface normal vector, and $\delta$ is the Dirac delta function with a unit of 1/length and maximum at surfaces [26]. Bound charges arise from splay deformations within the polarization field and are described by: $\rho_B = -\nabla \cdot \mathbf{P} = -P_0 \nabla \cdot \mathbf{n}$ [13]. Consequently, electrostatic interactions penalize both splay deformations and the tilting of polarization at the surfaces. The precise influence of these electrostatic interactions on polar ordering remains an intriguing and incompletely understood facet of ferroelectric nematic behavior and an active area of studies [10,19,26,34,35].

Here, we explore the polar order of ferroelectric liquid crystal RM734 under ionic doping. We reveal that confinement between glass plates coated with ionic polymers facilitates a stable modulated phase between the nematic and ferroelectric nematic phases. Optical microscopy reveals thickness-dependent periodicity, a sinusoidal variation of the orientation angle, and periodic variations in phase retardance and image intensity—key optical features consistent with theoretically predicted double-splay nematics. In this phase, splay deformations alternate along two orthogonal directions, oriented at 45° to the cell plane to ensure surface-parallel alignment of both polarization and director. We further show that only cationic polymer coatings effectively promote this double-splay nematic configuration, highlighting the importance of both charge and size of the dopant molecules. Finally, we examine the topological structures within the temperature range between the nematic and ferroelectric nematic phases. Upon cooling from temperatures above 190 °C, the liquid crystal progresses through the I, N, M, and $N_F$ phases, each exhibiting distinctive optical textures under polarized optical microscopy [Figs. 1(b-e) and Video S3 in Supplemental Material]. The corresponding transition temperatures are: $T_{I-N}$=184.0°C, $T_{N-M}$=128.0°C and $T_{M-NF}$=127.8°C.

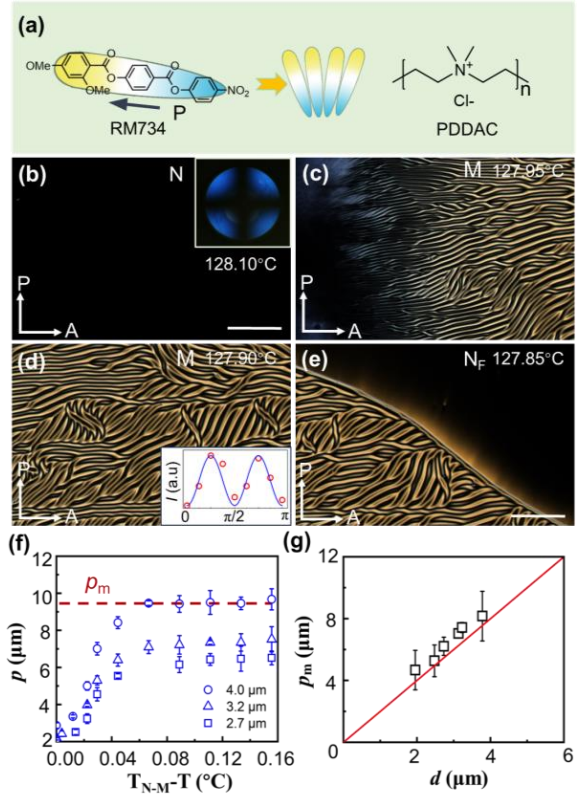

At ~127.85 °C, the modulated phase, characterized by periodic bright and dark stripe domains, emerges from the homeotropically aligned nematic phase. The transition is continuous, as evidenced by the diffuse interfaces between the phases and the gradual increase in modulation wavelength with distance from the interface [Figs. 1(c-d)]. At ~127.95 °C, the $N_F$ phase emerges from the modulated phase. This transition is first order as indicated by the sharp interface between the M and $N_F$ phases [Fig. 1(e)].

To investigate the temperature dependence of the modulation wavelength while minimizing the influence of minor temperature gradients in our hot stage, we focused on optical textures within a small viewing area (approximately 20 μm) at a fixed position. For a given cell thickness, the modulation wavelength ($p$) initially increases with decreasing confined double-splay nematic phase, revealing their distinct three-dimensional polarization characteristics.

**FIG. 1** Materials and the modulated phase. (a) Structures of the RM734 molecule and the cationic polymer PDDAC. (b-e) Polarizing optical microscopic images of the liquid crystal going through the nematic, modulated and ferroelectric nematic phases when temperature is lowered, in a cell of 3.4 μm thickness. The inset in (d) presents measured optical intensity at an arbitrary position versus the orientation angle with the polarizer. The axis marked with P and A represent the polarization direction of polarizer and analyzer, respectively. (f) Measured modulation wavelength $p$ versus temperature deviation from the onset temperature $T_{N-M}$. (g) Saturated modulation wavelength $p_m$ versus cell thickness. The red solid line represents $p_m = 2d$. Scale bars are 50 μm.

As shown in Fig. 1(a), the RM734 molecule possesses a pear shape with an electric dipole moment of approximately 9.8 Debye, pointing towards its larger end [4]. This liquid crystal exhibits a well-defined phase sequence : isotropic (I), nematic (N), intermediate phase ($N_X$), ferroelectric nematic ($N_F$), and crystalline (C), with the following transition temperatures: $T_{I-N}$= 184 °C, $T_{N-NX}$ = 132 °C, $T_{NX-NF}$= 131 °C, and $T_{NF-C}$ = 84 °C [41]. The intermediate $N_X$ phase exhibits modulated textures (Fig. S1, Video S1 and Video S2 Supplemental Material). However, its instability precludes structural characterization, unlike in ion-doped samples. Therefore, we adopt $N_X$ to represent the ambiguity surrounding its precise structure.

When RM734 is confined between glass plates coated with a thin film of the cationic polymer PDDAC [poly-diallyldimethylammonium chloride, depicted in Fig. 1a], a stable modulated (M) phase emerges within a narrow temperature, then plateaus at approximately 0.06 °C below the onset temperature $T_{N-M}$ of the modulated phase [Fig. 1(f)]. The saturated (maximal) modulation wavelength, $p_m$, exhibits a linear dependence on cell thickness ($d$), described by the relationship $p_m = 2d$ [Fig. 1(g)].

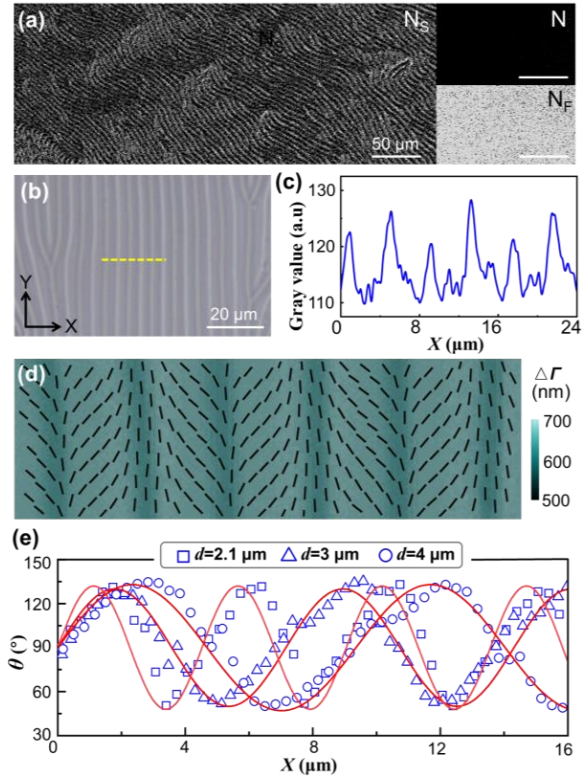

**FIG. 2** Polar order and optical microscopic analyses. (a) SHG images of the modulated, the nematic and ferroelectric nematic phases. (b) Bright field optical microscopic image of the modulated phase. (c) Representative intensity profile in the direction perpendicular to the stipes in (b). (d) Measured directors overlayed on the phase retardance. The scale bar is 5 μm. (e) Orientation angles along the X-axis measured at different cell thicknesses. The red lines are fittings with sinusoidal function $\theta(x) = \theta_0 \sin(kX)$.

We used second-harmonic generation (SHG) imaging to probe the polar order of the modulated (M) and ferroelectric nematic ($N_F$) liquid crystal phases. SHG images were acquired using a circularly polarized excitation laser at a wavelength of 800 nm, with signal detection at 400 nm. In the modulated phase, these images reveal periodic intensity modulation, while the ferroelectric nematic ($N_F$) phase exhibits uniform intensity [Fig. 2(a)]. Given that the SHG effect occurs only in materials lacking inversion symmetry, and that its intensity is directly proportional to the remnant polarization [42], these observations demonstrate a spatially varying polar order within the modulated phase.

Bright-field optical microscopy of the modulated phase reveals stripe domains separated by parallel bright lines [Fig. 2(b)]. A subtle yet key observation is the periodic alternation in brightness of these bright lines [Fig. 2(c)].

Cross-polarized optical microscopy demonstrates that the light intensity can be extinguished completely via sample rotation, indicating the absence of director twist across the cell [inset in Fig. 1(d)]. This allows us to measure the director field $\mathbf{n}(x, y)$ (i.e., the slow optical axis) and the retardance $\Delta\Gamma$ between optical waves polarized parallel and perpendicular to the director $\mathbf{n}$, using PolScope [43]. The measured director field reveals periodic splay deformations with alternating splay directions, while the retardance exhibits a small but discernible modulation [Fig. 2(d)]. We verified that regions of maximal splay correspond to the bright lines observed in bright-field images.

The orientation angles of the director relative to the stripe direction (Y-axis) exhibit a periodic variation along the perpendicular direction (X-axis) [Fig. 2(b)]. This variation is well-described by a sinusoidal function, with the fitted amplitudes remaining independent of cell thickness [Fig. 2(e)].

This sinusoidal director field appears to be consistent with the single splay nematic phase where the director field is described by: $\mathbf{n}(x) = [\sin\theta(x), 0, \cos\theta(x)]$ with $\theta(x) = \theta_0 \sin(kx)$ [31], where $\theta_0$ is the splay angle amplitude. However, this one-dimensional director configuration cannot account for three distinctive features of the M phase (see session 3 in supplemental material): the maximal modulation wavelength being twice of the cell thickness [Fig. 1(g)], the periodic spatial variation in retardance [Fig. 2(d)], the alternating brightness of the splay regions under bright-field microscopy [Figs. 2(b-c)]. These features, however, naturally arise from the double splay nematic phase, as explained below.

The double splay nematic phase comprises two types of domains arranged in a two-dimensional square lattice [Fig. 3(a)] [22,31]. Within each domain type, the polar order parameter $\mathbf{P}(x, y) = P_0(x, y)\mathbf{n}(x, y)$ and the splay vector $\mathbf{n}(\nabla \cdot \mathbf{n})$ are aligned in the same direction, while pointing in opposite directions in different domain types [Fig. 3(a)]. Both the polar order and splay vanish at the interfaces between these opposing domains[31]. The director $\mathbf{n}$ can be expressed as [Fig. 3(a)] [31]:

$\mathbf{n}(x, y) = [\theta_0 \sin kx \cos ky, \theta_0 \cos kx \sin ky, 1]/n_0$ (1)

where $n_0$ is a factor ensuring that $|\mathbf{n}| = 1$. The amplitude of the polar order parameter is approximately: $P_0(x, y) \approx 2p_0 \cos kx \cos ky$ [31]. Within the (110) planes defined by $y = x + \pi m/k$ where $m$ is an integer, $n_x = n_y$ and the director is parallel to these planes [dashed lines in Fig. 3(a)].

Under confinement between parallel glass plates, the director (polarization) at the surfaces must align parallel to the cell plane to minimize surface charge accumulation. The shortest period of the double splay nematic capable of satisfying this boundary condition is that bounded by two neighboring (110) planes, as indicated by the dashed lines in Figs. 3(a-b).

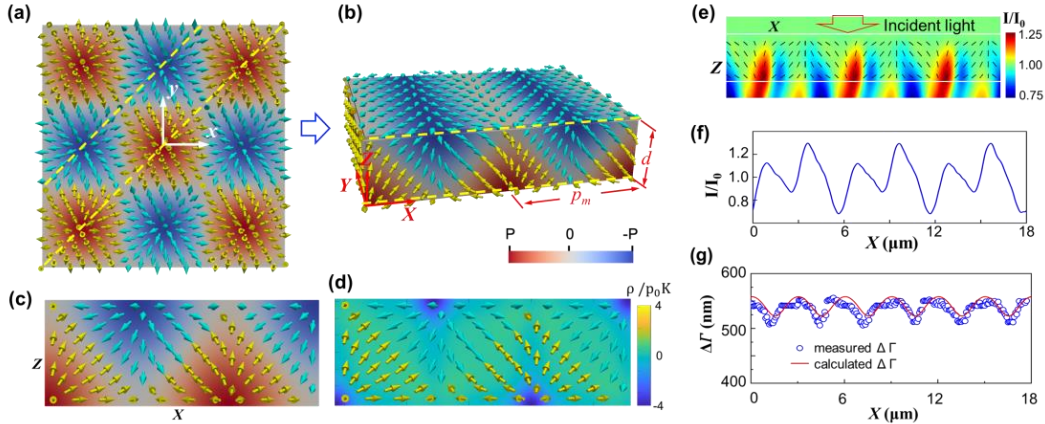

**FIG. 3** Structures of double splay nematic phase. (a) Polarization field of the double splay nematic in free space. (b) Polarization field of the double splay nematic confined between two flat surfaces. (c) Polarization field in XZ plane. Arrows indicate polarization direction, while background colors represent polarization amplitude, as shown by the color scale in (b). (d) Distribution of bound charge density in the XZ plane, calculated in units of $p_0K$. (e) FDTD-calculated optical field distribution (XZ plane) for circularly polarized incident light (400-780 nm). (f) Light intensity at the bottom liquid crystal-substrate interface. (g) Calculated retardance (solid red line) compared to experimental data (data points).

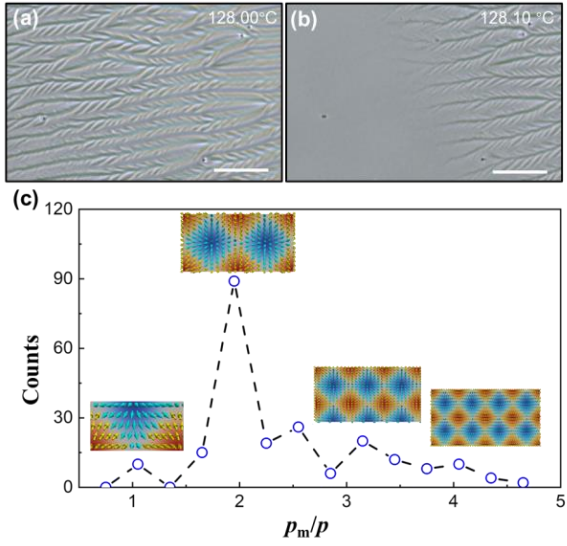

**FIG. 4** Transition from the double splay nematic to nematic. (a-b) Representative bright field images of evolving stripe domains upon heating. The scale bars represent 20 μm. (c) Counted numbers of sub-domains at different spatial modulation wavelength ($p$). The inset images are the polarization fields in the XZ plane corresponding to $p_m/p = 1, 2, 3, 4$.

The director field of this configuration in XYZ coordinates can be obtained by applying a 45° rotation about the z-axis, followed by a 90° rotation about the x-axis, represented by the rotation matrix $R = R_z(\pi/4)R_x(\pi/2)$. The director field can be obtained via $\mathbf{n}(X,Y,Z) = R\mathbf{n}(x,y,z)$ as:

$$\mathbf{n}(X,Z) = [\theta_0 \sin(KX), 1, -\theta_0 \sin(KZ)]/n_0 \quad (2)$$

where $K = k/\sqrt{2}$. The polar order parameter is given by: $P_0(X,Z) = 2p_0\cos[K(X-Z)]\cos[K(X+Z)]$. This leads to the maximal modulation wavelength being twice of the cell thickness $p_m = 2d$ [Fig. 3(b)], and thus $K = 2\pi/p_m = \pi/d$.

Figs. 3(b-c) illustrate the polar order parameter $\mathbf{P}(X,Z)$ in the XY and ZY planes. Notably, the polarizations at the two confining surfaces are antiparallel. Fig. 3(d) presents the spatial distribution of bound charge density $\rho_B = -\nabla \cdot \mathbf{P}$, in XZ plane and plotted in units of $p_0 K$.

We calculated optical field distributions for such confined double splay nematics by numerically solving the Maxwell equations using finite-difference time-domain (FDTD) method. The FDTD method, widely used for calculating optical properties of anisotropic media like liquid crystals, provides a robust approach to modeling light propagation in these complex systems [44,45]. Further details of the calculations are included in the Supplemental Material.

While the interaction of light with the modulated liquid crystal is complex [Fig. 3(e)], the calculated intensity profile at the interface between the liquid crystal and substrate closely resembles that observed in bright-field microscopy [Fig. 3(f) and Fig. 2(c)]. This can be attributed to the alternating triangular domains acting as miniature lenses, focusing light into lines. The top and bottom triangles give rise to different light intensities due to variations in their focusing power. In contrast, the calculated intensity distribution at the liquid crystal interface for single splay nematics does not exhibit this alternating intensity variation (Fig. S2).

To understand the PolScope results, we look at the director field $\mathbf{n}(X,Z)$, which represents the principle optical axis at position (X, Y, Z). The director's polar angle with respect to the Z-axis is given by: $\varphi(X,Z) = \arctan\left[\sqrt{1/\theta_0^2 + \sin^2 KX}/\sin KZ\right]$, and its azimuthal angle $\phi$ with respect to X-axis is: $\phi(X) = \arctan[1/\theta_0 \sin(KX)]$. The azimuthal angle in independent of $Z$, indicating the absence of twist across the cell. For light propagating along the Z-direction, the refractive index is $n_o$ when the polarization is normal to the plane defined by the Z-axis and the vector $\mathbf{n}$. When the polarization is parallel to this plane, the effective refractive index is: $n_{\text{eff}} = (\cos^2\varphi/n_o^2 + \sin^2\varphi/n_e^2)^{-1/2}$. The retardance between these two orthogonally polarized optical waves is then calculated as: $\Delta\Gamma = \int_0^d (n_{\text{eff}} - n_0)dZ$. Using $n_e=1.72$ and $n_o=1.51$ (from reference [46]) and the experimental cell thickness $d$, we numerically calculated $\Delta\Gamma$ for varying X. This calculation reproduced the periodic variation of $\Delta\Gamma$ observed in the PolScope measurements, yielding good agreement with the experiments [Fig. 3(g)].

Another evidence supporting this confined double splay configuration comes by studying the transition from the double splay nematic to the N phase. When heated at a rate of 0.05°C/min, new stripes with smaller modulation wavelengths emerge within the stripe domains, with a tilt angle between them [Fig. 4(a)]. At ~128.10°C, these sub-stripes smear out to become a homogeneous nematic (N) phase, with a continuous phase boundary [Fig. 4(b)]. We measured the separation ($p$) between neighboring sub-stripe domains and counted the number of sub-stripe pairs with similar modulation wavelengths [Fig. 4(c)]. The maximal occurrence appears at $p_m/p = 1, 2, 3, 4$. This is consistent with the confined double splay nematic model, namely the newly formed stripe sub-domains consist of diagonal double splay nematic layers bounded by (110) planes, as schematically illustrated in the insets of Fig. 4(c).

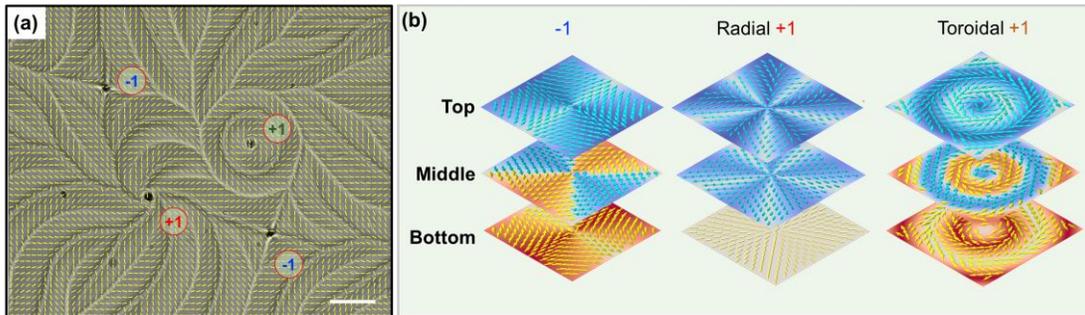

**FIG. 5** Topological defects in confined double splay nematics. (a) Exemplary measured director field overlayed on bright field optical image, where -1, toroidal +1 and radial +1 topological defects are indicated. Scale bar is 10 μm. (b) Constructed polarization fields of -1, toroidal +1 and radial +1 topological defects.

The RS theory provided mathematical expressions for the parameters, $\theta_0$, $p_0$ and $k$ [31]. Near the critical temperature of the N-N$_S$ transition, $\theta_0$ is related to the temperature and elastic constants by: $\theta_0 = 2^{3/2} K_{11} \delta \mu^{1/2}/(3 K_{33})^{1/2} \lambda$, where $\mu = \mu'(T - T_0)$. Given that $T - T_c$ is less than 0.2 °C, we can estimate that the high order term is negligible. By taking $K_{11} = 1$ pN, $K_{33} = 1$ pN, $\mu = -1 \times 10^3$ Jm/C$^2$, $\lambda = 10^{-4}$ V [14], we can estimate that $\theta_0 \sim 60°$, which is reasonably close to the experimental data ~80°.

The RS theory shows that the modulation wavelength decreases rapidly and reaches a minimum right below the N-N$_s$ transition temperature and then go up again when the temperature is lowered. We only observed growths of the modulation wavelengths with decreasing temperature [Fig. 1(f)], that is in good agreement with the second region of the theoretical prediction.

Ionic coatings on the confining surfaces play a critical role in the formation of the double splay nematic phase. We evaluated cationic polymers, anionic polymers, and small ionic molecules (Fig. S3 and Table S1). Robust double splay nematic ordering was observed exclusively with cationic polymeric coatings; anionic polymers failed to induce any modulated phase. We noted an intermediate modulated phase with the small ionic surfactant CTAB, but its textures and topological defects are quite different, and its structure requires further investigation to confirm its identity as a double splay nematic. These results indicate that both the charge and size of the dopants significantly influence the formation of the double splay nematic phase.

The efficacy of cationic polymeric coatings in promoting double splay nematic order under confinement can be attributed to their unique ability to address both surface and bulk bound charges. The double splay nematic configuration under confinement results in primarily negative bound charges at the surfaces and positive bound charges in the bulk [Fig. 3(d)]. The cationic polymers remain at the surfaces to neutralize these negative surface charges, while dissociated negatively charged ions from the polymers are mobile and can effectively compensate for positive bulk bound charges by reducing the Debye screening length. This interaction between the cationic coatings and the bound charge distribution enables the stabilization of the double splay nematic phase.

The double splay nematic phase exhibits intriguing topological defects (Fig. S4), emerging from the N phase due to the degenerate orientations of the stripe domains at different areas (Video. S4). Upon transitioning into the double splay phase, these topological defects may annihilate with each other. We have observed three distinct types: -1 defects with quartic symmetry, +1 defects with radial splay stripes, and +1 defects with toroidal splay [Fig. 5(a)]. Based on our understanding of the confined double splay nematic, we reconstructed the polarization fields of these defects using optical microscopy measurements.

The -1 defect features a singular defect center where four splay stripes converge [Fig. 5(b)]. Bright-field optical imaging reveals alternating brightness in these four maximal splay regions, indicating that they reside on opposite surfaces (Fig. S5). In contrast, the +1 defect with radial splays displays five splay stripes radiating outward from the center, with both splay and polarization vectors oriented radially [Fig. 5(a)]. These stripes are located on a single surface, as evidenced by the uniform brightness of the five splay regions under bright-field imaging [Fig. 5(b)]. For the +1 defect with toroidal splay, the director field exhibits rotational symmetry. Bright-field imaging shows a dimmer inner splay region compared to the outer splay region [Fig. 5(a)], indicating that these two splay regions originate from different confining surfaces. Consequently, the rotation directions of the polarization vectors are opposite at the two surfaces [Fig. 5(b)].

Another ferroelectric nematic material is the DIO, which is known to exhibit an intermediate antiferroelectric phase, called the SmZ$_A$ phase[37]. We explored the effects of ionic doping on this intermediate phase and found that the optical textures are quite complex (Fig. S7 and Fig. S8), drastically different from those modulated textures observed with RM734. Given the ~18 nm modulation periodicity of the intermediate phase in pure DIO, we believe that any modulated phases in ion-doped DIO would be beyond the resolution of our optical imaging techniques. One interesting comment of the RS paper is that the modulated phases with small periodicity are likely single splay [31].

In conclusion, we have shown that double splay nematic order can be enabled to form in confined polar fluids by modifying surface charges and electrical screening via cationic polymer coatings, where splay deformations alternate between two orthogonal directions oriented at 45° to the cell plane, aligning both the director and polarization parallel to the cell surfaces. These findings not only confirm the existence of the theoretically predicted double splay order but also provide an approach to control the formation, orientation and periodicity of the double splay nematics by using ionic surface coating and tailored confining geometries. Given its strong nonlinear optical properties, such control holds significant potential for applications in nonlinear and quantum optics. The experimental finding of this double splay nematic ordering will inspire further research into both novel ordering induced by flexoelectricity and electrostatic interactions and potential applications of their unique properties.

Note added—A research work suggesting a double splay nematic in ionic-liquid-doped RM734 appeared recently [47]. Their optical textures and topological defects are quite similar to those in our experiments with ionic surfactant CTAB doping [Figs. S3(b)–S3(d)].

We acknowledge the financial support by the National Key Research and Development Program of China under grant 2022YFA1405000, the National Natural Science Foundation of China under grants 12174177 and 62105141, and the Guangdong Basic and Applied Basic Research Foundation under grant 2024B1515040023. We extend our gratitude to J. V. Selinger and R. B. Selinger for their valuable discussions and to Zening Hong for his technical assistance.

---

*Correspondence to: weiqh@sustech.edu.cn